\documentclass[twocolumn]{aastex63}

\usepackage[utf8]{inputenc}
\usepackage{microtype}
\usepackage{graphicx}
\usepackage{color}
\usepackage{float, times}
\usepackage{amsmath, amssymb, amsfonts, amssymb}
\usepackage[nice]{nicefrac}
\usepackage[T1]{fontenc}
\usepackage{url}
\usepackage[colorlinks=true]{hyperref}
\usepackage{natbib}

\submitjournal{ApJ}

\begin{document}
\title{Investigation of surface effects of simple flux tubes using numerical simulations}
\author{M. Waidele}
\author{M. Roth}
\affil{Leibniz-Institut für Sonnenphysik, D-79104, Germany}
\email{waidele@leibniz-kis.de}

\begin{abstract}
	We use the SPARC code for MHD simulations with monolithic flux tubes of varying subsurface topology. 
	Our studies involve the interactions of waves caused by a single source with subsurface magnetic fields.
	Mode conversion causing acoustic power to trickle downwards along the flux tube has been described before and can be visualized in our simulations.
	We show that this downward propagation causes the flux tube to act as an isolated source, creating a characteristic surface wavefield.
	Measuring this wavefield at the surface reveals subsurface properties of the magnetic field topology.
	Using time distance helioseismology, we demonstrate how to detect such a flux tube signal based on a group travel-time delay of $\varDelta t = 282.6$ sec due to the wave packet spending time subsurface as a slow mode wave.
	Although the amplitude is small and generally superimposed by the full wave field, it can be detected if assumptions about $\varDelta t$ are made.
	We demonstrate this for a simulation with solar like sources.
	This kind of study has the potential to reveal subsurface information of sunspots based on the analysis of a surface signal.
\end{abstract}

\section{Introduction}
\label{sec:Intro}
	Sunspots play an important role in understanding the dynamical nature of the solar magnetic field. 
	Although their surface appearance has been observed for over four centuries, little is known about the subsurface structure. 
	They are known to strongly influence solar acoustic modes and there is a variety of possible interactions of the magnetic field with waves in its proximity.
	Sunspot seismology is the study of these waves and their interactions with the goal of understanding more about subsurface properties, such as magnetic field configurations, mass flows and thermal structures.
	
	Earlier studies revealed that most techniques of helioseismology break down in the presence of strong magnetic fields \citep{2010SoPh..267....1M, 2009SSRv..144..249G} and can therefore not be used in sunspot seismology. 
	Since there are additional issues with velocity measurements within sunspots, such as the change of height due to the Wilson Depression, atomic lines being affected by magnetic fields and the suppression of oscillations \citep{1990ApJ...354..372B}, the analysis of data can be difficult. 
	Often times the study of isolated and simplified effects therefore rely on MHD simulations. 
	Efforts have been made to find discernible surface signatures caused by flux tubes with different subsurface structures. 
	Examples include investigations of acoustic Halos around sunspots \citep{2013SoPh..287..107R, 2016ApJ...817...45R} and scattered wavefields after interaction with magnetic fields \citep{2011SoPh..268..429Z}. 
	\cite{2013A&A...558A.130S} studied changes to observable travel times caused by abnormalities in the subsurface structure of a flux tube.
	
	An important physical quantity to consider when wave fields are studied is the $c_\text{A} = c_\text{s}$-layer where the sound-speed $c_\text{s}$ is equal to the Alfvèn-speed $c_\text{A}$ \citep{2002ApJ...564..508R, 2007AN....328..286C}. 
	At this layer mode conversion is most prominent. 
	Mode conversion is fundamental in understanding wave behavior in and around sunspots. 
	Thus the (well established) p-mode absorption for example close to active regions \citep{1987ApJ...319L..27B, 1988ApJ...335.1015B} could eventually be contributed to conversion of waves in slow mode waves traveling along the flux tube \citep{1997ApJ...486L..67C}. 
	Also most interpretations nowadays of the formation of acoustic halos around active regions  include mode conversion in some layer within the solar atmosphere \citep{2008ApJ...680.1457H, 2012ApJ...746...68K, 2012A&A...538A..79N}.
	
	When considering MHD simulations a number of simplifications and limitations need to be applied, so the computational expense remains reasonable. 
	Especially when only atmospheric effects on wavefields are studied, it is justifiable to neglect radiative transfer and therefore convection and granulation. 
	An appropriate code for solving the MHD equations for seismic propagation is the SPARC code, developed by \cite{2007PhDT........24H, 2007AN....328..319H}. 
	It has been used extensively in the past. 
	\cite{2016ApJ...817...45R} used it to analyse the effect of the Alvèn-limiter on the formation of the before mentioned acoustic halos. 
	\cite{2009A&A...501..735S, 2015ApJ...807...20P, 2015ApJ...801...27R} carried out single source excitation simulations to study how acoustic power is (distributed) along and around flux tubes.
	
	In this work we first visualize the aforementioned slow mode waves traveling downwards into the interior of the simulations domain. 
	This downward propagation will eventually turn acoustic waves back up to the surface, where they can be measured. 
	Assuming that these waves carry information, amongst other things, about the subsurface extent of the $c_\text{A} = c_\text{s}$-layer, a rough image of magnetic field configuration with height can be obtained.

	The simulation set up is described in section \hyperref[sec:2]{\ref{sec:2}}, including the artificial atmosphere (section \hyperref[sec:2.1]{\ref{sec:2.1}}) plus the flux tube model (section \hyperref[sec:2.2]{\ref{sec:2.2}}). 
	In section \hyperref[sec:3]{\ref{sec:3}} we analyse surface effects of the flux tube in the very simple scenario of one isolated source. 
	Concerning more realistic simulations, we show in section \hyperref[sec:4]{\ref{sec:4}} how to potentially detect such surface effects in real data.

\section{Simulation set up}
	\label{sec:2}
	SPARC is a code that can be used to compute the interactions of waves with magnetic flux tubes, sound-speed and damping perturbations, and study the wave field in the presence of multiple/single sources or anomalies thereof. The linearized MHD and Euler equations in 3D Cartesian geometry are solved. The derivatives are computed using sixth-order compact finite differences (in all three directions) or FFTs in the horizontal directions and an optimized second-order RK time stepping scheme is implemented \citep{SMH07}.

	As the goal of this work is to find information about subsurface structures at the surface carried by propagating waves, a solar like background stratification including a magnetic field described by flux tube model needs to be set up. 
	
	Our simulation domain is enclosed in a box with $256\times 256\times 300$ grid points. The dimensions are $373.76\times 373.76\times 40$ Mm$^3$. $x$ and $y$ dimensions are required to be chosen such that typical (solar like) acoustic wavelengths are resolved. Thus, they were set to be comparable to available data sets of solar surface velocities, such as HMI (link to HMI). For the maximum depth, one has to usually find a trade-off between resolution (i.e. computational expense) and the existence of modes with deep turning point within the simulation domain. $40$ Mm is hereby a reasonable choice.
	
	When dealing with isolated sources as in section \hyperref[sec:3]{\ref{sec:3}}, the boundaries in each direction included a PML-layer to absorb outgoing waves as efficiently (with as little reflection) as possible \citep{2010A&A...522A..87H}. This is reasonable for this particular scenario, since we consider single wave packets crossing each grid point only once. For the stochastic excitations considered in section \hyperref[sec:4]{\ref{sec:4}} the horizontal boundaries were chosen to be periodic, in order to keep the simulation as realistic as possible.
	
	For every simulation run, the vertical velocities $v_z$ at the surface ($z = 0$ Mm) are written out and stored at a cadence of $\varDelta t = 45$ sec (again, in order to resolve solar like acoustic frequencies and to be similar to HMI data sets). In the following, vertical surface velocities stemming from simulations with the fully magnetized atmosphere are labeled as $v_\text{magn}$. In addition wave fields for quiet runs yielding $v_\text{quiet}$ were obtained. 
	Quiet in this case means an atmosphere without any magnetic field (effectively 1D).
	The idea is that the difference of the velocities $v_\text{diff} = v_\text{magn} - v_\text{quiet}$ is basically a noise subtraction (where $v_\text{quiet}$ is identified as noise), highlighting remnants of the full wave field influenced by the presence of the magnetic field. 
	These remnants are mostly waves caused by gradients in pressure, density and sound-speed due to modifications of the background (by the magnetic flux tube), but also waves originating from mode-conversion. 
	This difference signal $v_\text{diff}$ does not represent the scattered wave and might not have a trivial physical meaning, but it still lets us investigate the mode conversion within the flux tube (and any wave field that results from it) much easier than taking only the full wave field $v_\text{magn}$ into account.
	Observation of $v_\text{diff}$ in real data is generally not possible, however in section \hyperref[sec:4]{\ref{sec:4}} we present a method of measuring it, to some extent, indirectly.

	Additional simulation are set up, containing only the thermal perturbations of the fluxtube within the atmosphere, but not the magnetic field itself (thus yielding $v_\text{therm}$). We use these thermal only runs, as qualitative measure for the direct influence of the magnetic field, similar to \cite{2015ApJ...801...27R}. We refrain to use the difference $v_\text{diff}^\prime = v_\text{magn} - v_\text{therm}$ for the quantitative analysis, since there is no trivial relation between $v_\text{diff}^\prime$ and $v_\text{diff}$.

	\subsection{Stabilized background atmosphere}
		\label{sec:2.1}
		As background atmosphere, a slightly modified version as presented in \cite{2007PhDT........24H} of model S \citep{1996Sci...272.1286C} is used. The modifications include changes to pressure and density such that the Brunt-Väisälä frequency $N$ is always real valued ($N^2 \geq 0$), in order to maintain a convectively stable surface layer. These changes entail some unphysical effects within the simulation domain. For example overstable waves: These are essentially extremely slowly propagating g-mode waves, which do not exist in the Sun. They can however easily be ignored, due their comparatively low group speed and frequency.
		
		Note also that the code does not include radiative transfer. Due to the additional convective stability, simulations will not include any convection or granulation. This makes the resulting wave behavior less realistic, but also much more simple to analyze.

	\subsection{Monolithic self similar flux tube model}
		\label{sec:2.2}
\begin{figure}[htb]
			\centering
			\includegraphics[width=.57\textwidth]{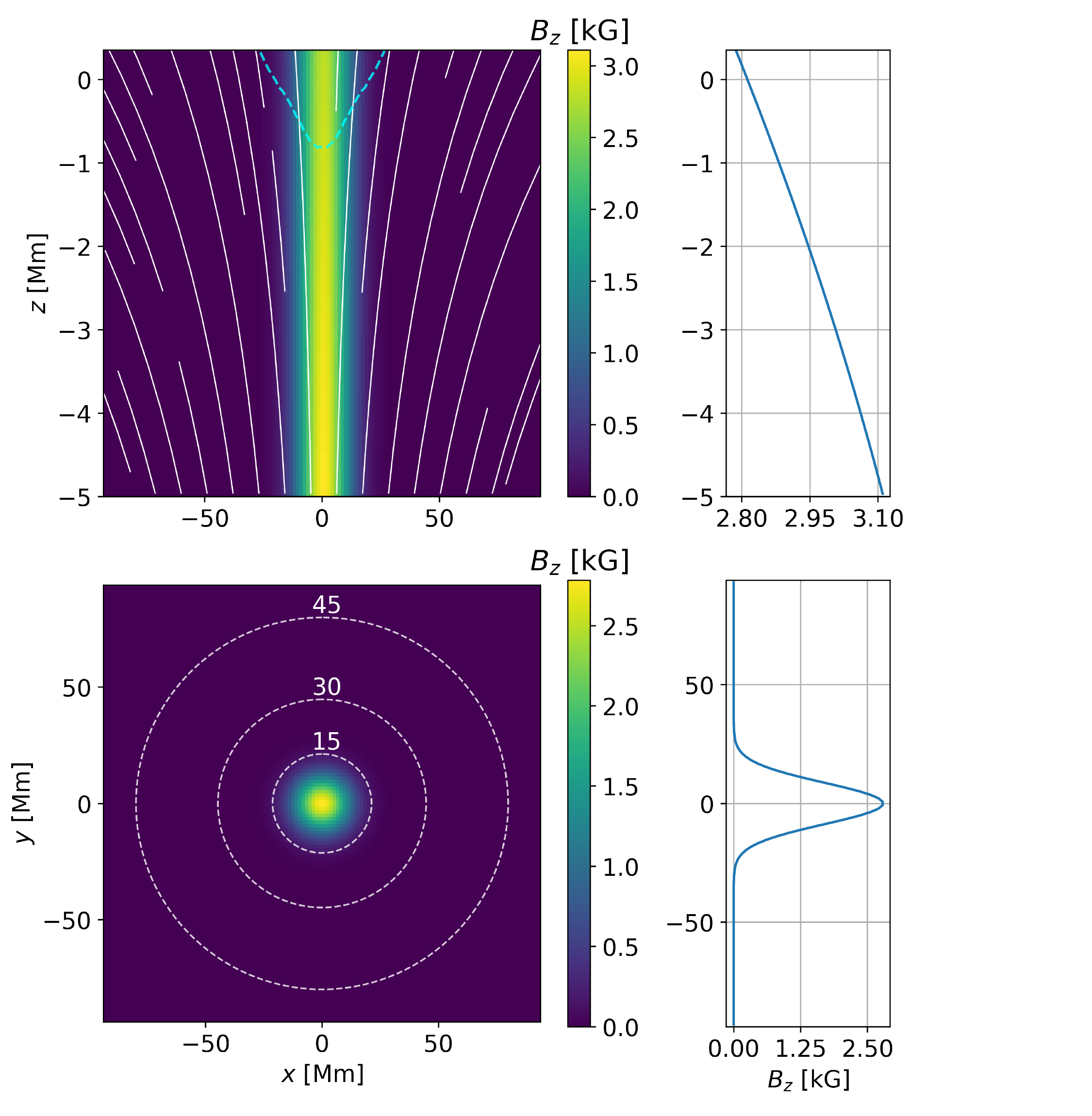}
			\caption{
				Monolithic self similar flux tube model used in the simulations (Further denoted as model 1). The surface peak strength of the vertical magnetic field is $2.8$ kG. A vertical slice ($x$-$z$-plane) is shown in the upper panel and a horizontal slice ($x$-$y$-plane) in the lower panel. White lines show the inclination, the dashed cyan line depicts the layer where $c_\text{A} = c_\text{s}$. The side panels are distribution taken from a 1D line through the center of the according image. Only a fraction of the full simulation domain is shown, depicting the essential properties of the magnetic field. \label{fig:1.1}
			}
		\end{figure}

		For our flux tube model we use a monolithic, self similar description as presented in \cite{1958IAUS....6..263S, 2008ApJ...680.1457H}. The basic structure can be seen in Figure \hyperref[fig:1.1]{\ref{fig:1.1}}. The demonstrated magnetic field distribution is further denoted as model 1. To deal with the quickly growing Alfvèn speed $c_\text{A}$ in the $z > 0$ layers (and especially within the flux tube), we introduced an Alvèn-limiter $c_\text{A}^\text{max} = 90$ km/s. Alfvèn-limiters should be set as high as possible, since it degrades the realism of the simulation further. Large values however make the computational expense large. $c_\text{A}^\text{max} = 90$ km/s is shown to be a good compromise in \cite{2016ApJ...817...45R}, and allows us to set the simulation time step to $\Delta_\text{ts} = 0.2$ sec. This work focuses more on the subsurface wave interactions, thus having a large $c_\text{A}^\text{max}$ is not crucial. $\Delta_\text{ts} = 0.2$ sec allows reasonable simulation wall times.
		
		\begin{figure*}[htb]
			\centering
			\includegraphics[width=1.\textwidth]{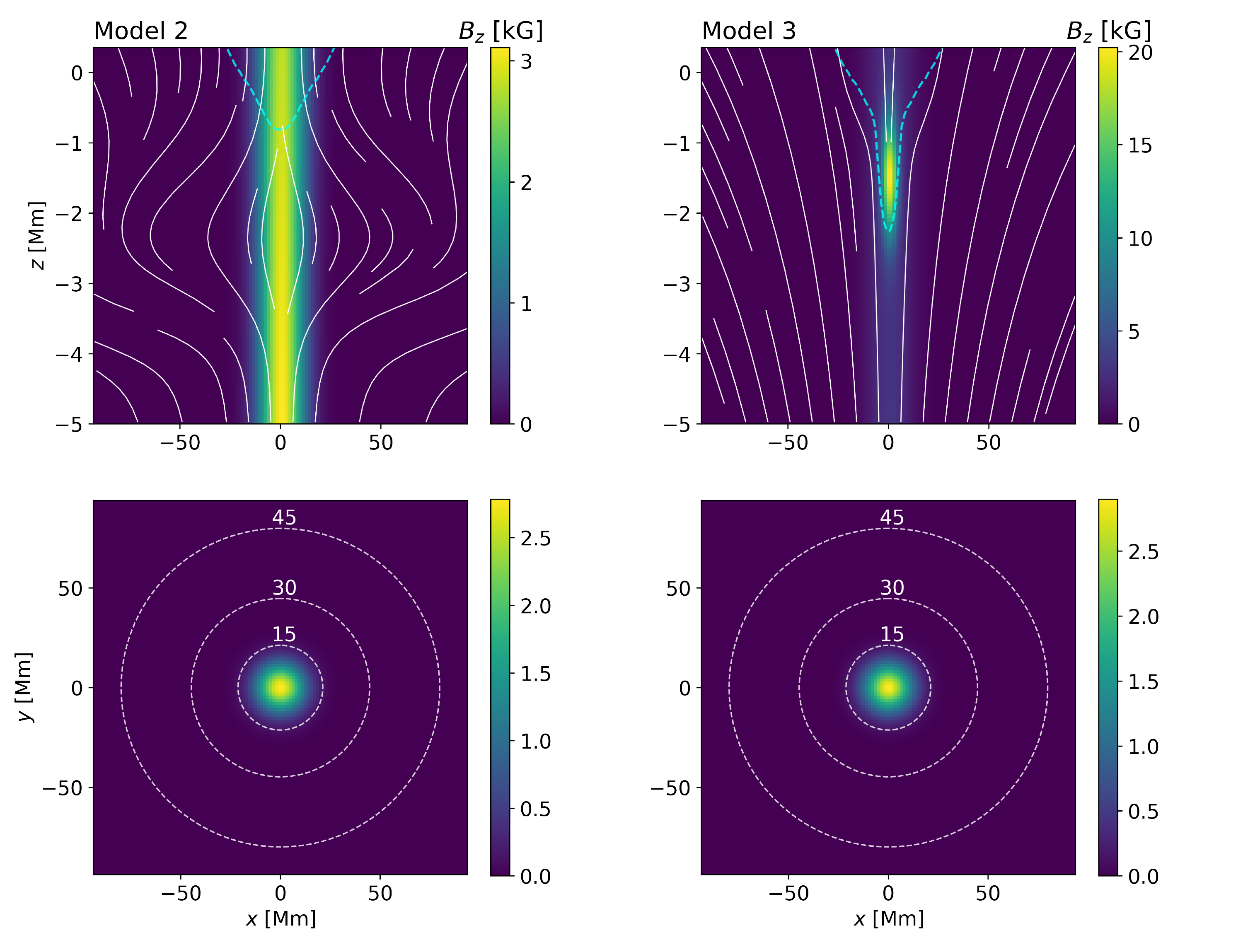}
			\caption{
				Fluxtube "toy" models 2 and 3. The surface peak strength of the vertical magnetic field is $2.8$ kG. A vertical slice is shown in the upper panels, a horizontal slice in the lower panels. White lines show the inclination, the cyan line depicts the layer where $c_\text{A} = c_\text{s}$. Note the vast changes of subsurface configurations compared to \hyperref[fig:1.1]{\ref{fig:1.1}}, but the very similar surface appearance. Not shown is the horizontal component of the magnetic field, causing model 2 to broaden at $z = -2$ Mm and model 3 to become more inclined at $z = -1.5$ Mm.\label{fig:1.2}
			}
		\end{figure*}
		
		With the method described in \cite{1958IAUS....6..263S}, the $c_\text{A} = c_\text{s}$-layer surfaces at $r(z=0) = 19.79$ Mm away from the center of the flux tube. This is similar to a rather large, but still realistic sunspot. For the purpose of this work, two more (not necessarily physical) flux tube models were constructed, labeled Model 2 and Model 3. Model 2, shown in Figure \hyperref[fig:1.2]{\ref{fig:1.2}}, left panels, is inspired by \cite{2013A&A...558A.130S}, including a broadening of the flux tube at a depth of $z = -2$ Mm. A broadening like this might be a consequence of convective motions, fanning out the field lines. Model 3, see Figure \hyperref[fig:1.2]{\ref{fig:1.2}}, right panels, was infused with a sudden increase of the vertical magnetic field $B_z$ at $z = -1.5$ Mm. This results in a stronger depression of the $c_\text{A} = c_\text{s}$-layer, affecting the subsurface mode conversion, while still preserving the wine-glass structure. These toy models exhibit approximately the same surface parameters (i.e. $B_z(z=0) \approx 2.8$ kG, $r(z=0) = 19.79$ Mm, etc.), which is a requirement in order to see if surface effects are affected only by subsurface properties.

		\subsection{Mode conversion and downward propagation}
			\label{sec:2.3}
			In order to keep the interactions at the flux tube boundary (that is $c_\text{A} = c_\text{s}$) as simple as possible, we employ a method similar to \cite{2009A&A...501..735S}. As mentioned before, only a single, quickly decaying oscillatory background displacement is used. It is quantified as:
			\begin{align}
				\nonumber v_z = \sin\left(\frac{2\pi t}{p_t}\right)\exp\left(-\frac{(t-t_0)^2}{\sigma_t^2}\right)\exp\left(-\frac{\lvert(\vec{x}-\vec{x_0})\rvert^2}{\lvert\vec{\sigma_x}\rvert^2}\right)
			\end{align}
			where $p_t$ is the oscillation period, $t_0$ the starting time, $\sigma_t$ the temporal width (i.e. length), $\vec{x_0}$ the location and $\vec{\sigma_x}$ the spatial width. We set $p_t = 302$ sec, equating to $\approx 3.31$ mHz. Also $t_0 = 200$ sec, $\sigma_t = 75$ sec (therefore quickly decaying), $\vec{x_0} = \left( -0.4, 150.0, 186.9 \right)$ Mm (origin of the $x$-$y$-axis at the corner of the box) and $\vec{\sigma_x} = \left( 0.6, 1.5, 1.5 \right)$ Mm.
			
\begin{figure*}[ht!]
				\centering
				\includegraphics[width=1.\textwidth]{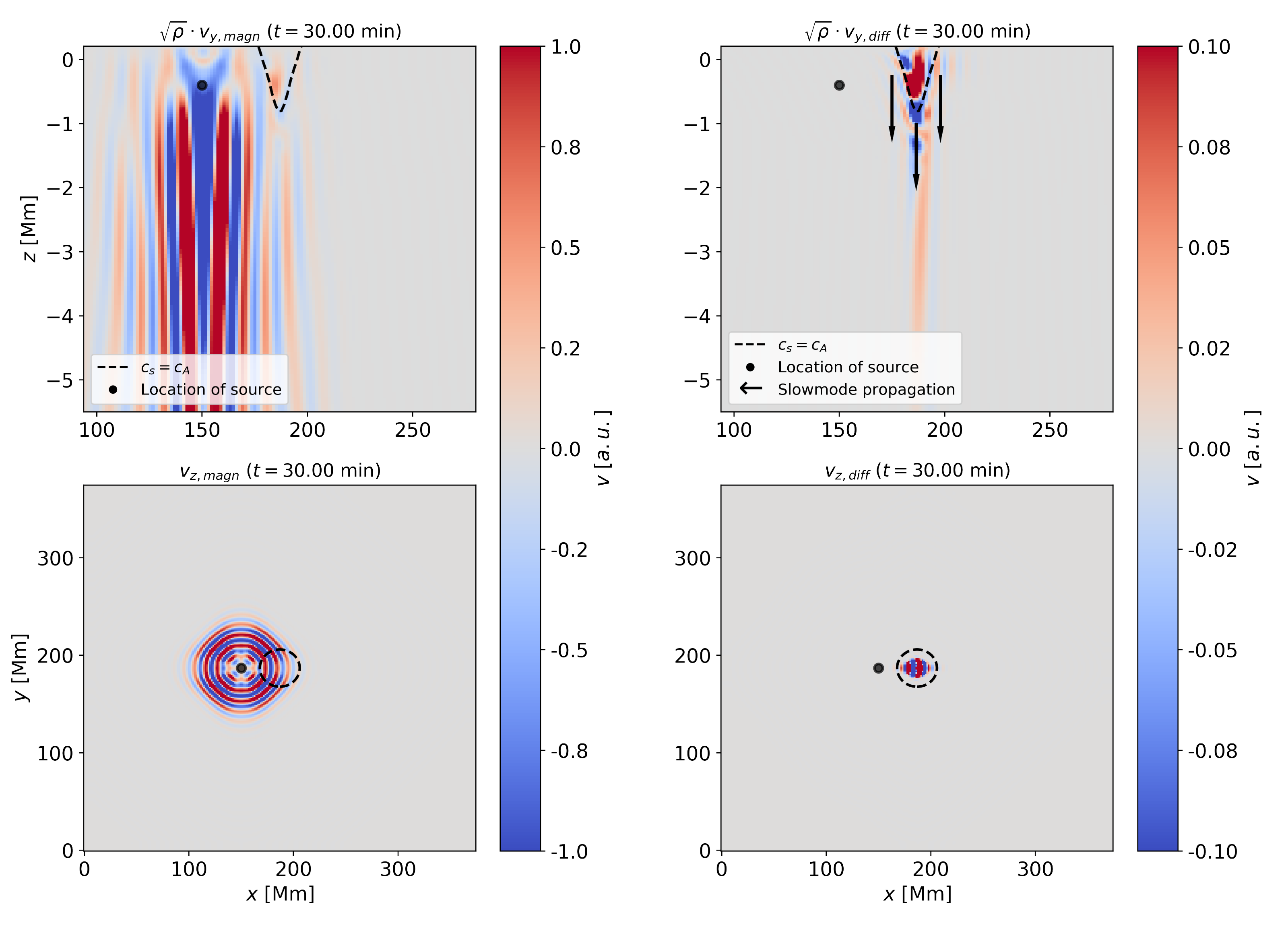}
				\caption{
					Wave propagation in the simulation box for vertical (top row) and horizontal (bottom row) cuts for $t=30$ min. The left columns show the full wave field $v_\text{magn}$, caused by a wave packet originating from the location of the source (black dot). The right columns show the instantaneous difference $v_\text{diff}$, only depicting waves that are caused by the presence of the magnetic field. Upper panels showing $v_y$ are scaled with $\sqrt{\rho}$. The $c_\text{A} = c_\text{s}$-layer is shown as black dashed line, black arrows indicate the predominant direction of wave propagation, after mode conversion at the $c_\text{A} = c_\text{s}$-layer. Note that the amplitude of $v_\text{diff}$ is about 10\% of the full wave field amplitude. \label{fig:2}
				}
			\end{figure*}

			This displacement will cause the propagation of a wave packet in every direction, simulated by SPARC. In Figure \hyperref[fig:2]{\ref{fig:2}} this propagation is visualized. The left panels show snapshots of the full wave field $v_\text{magn}$ where the upper panels show the $y$-component, and the lower panels the $z$-component. 
			The simulation of the wave field seen in the left-hand lower panel corresponds to the (theoretical) propagation of a single wave packet on the solar surface.
			For the chosen time $t = 30$ min, the separation of multiple skip branches (see sec. \hyperref[sec:3]{\ref{sec:3}}) already becomes visible. Note that there are some artificial features, such as absorption at the boundaries and amplitude distortion due to the Cartesian geometry, which is weak enough to be negligible. In the right panels, the instantaneous difference $v_\text{diff}$ is shown. 
			
\begin{figure*}[ht!]
				\centering
				\includegraphics[width=1.\textwidth]{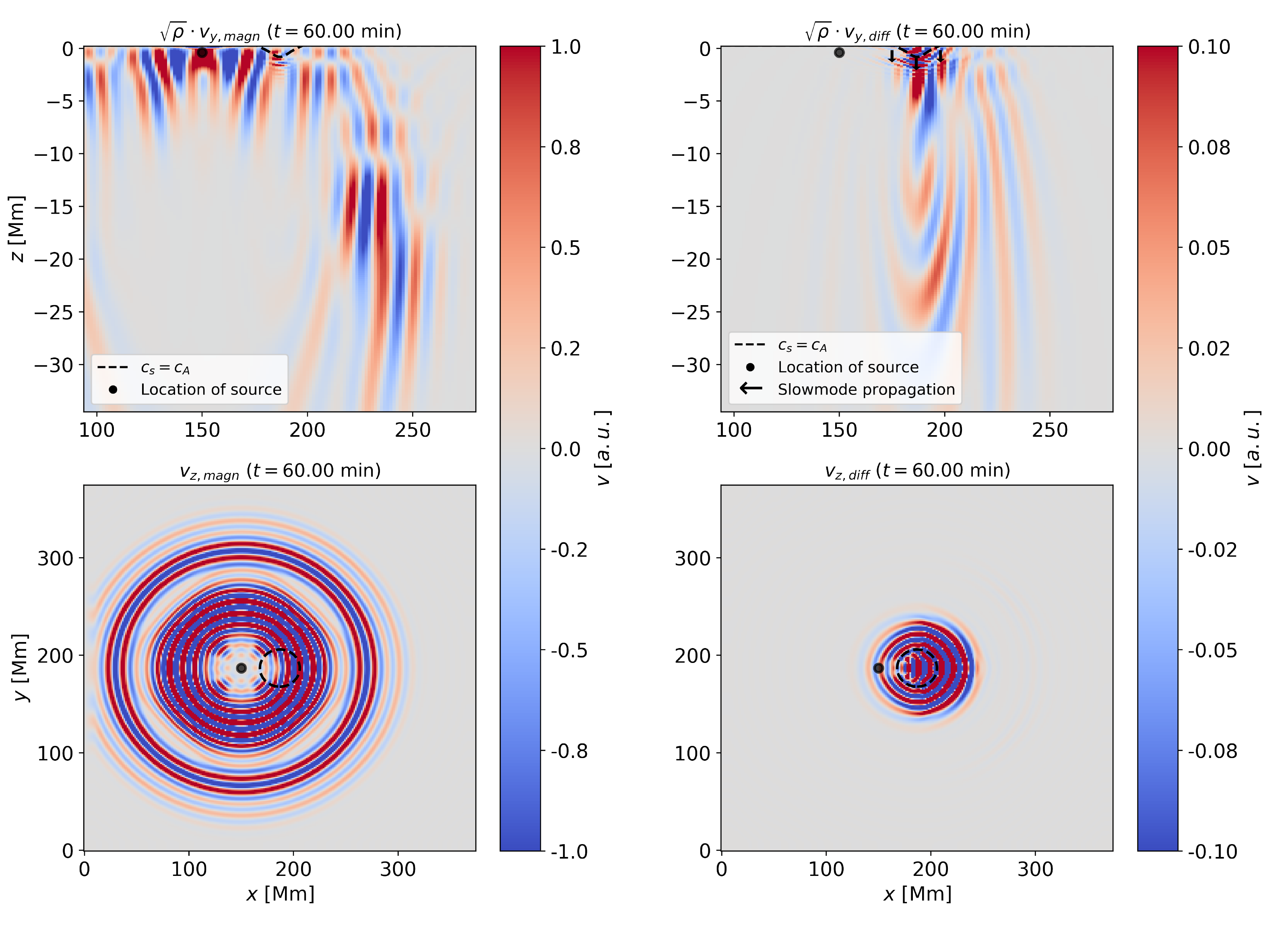}
				\caption{
					Same as \hyperref[fig:2]{\ref{fig:2}}, but for $t=60$ min. 
					The upper panels show the whole depth of the simulation domain (other than Fig. \hyperref[fig:3]{\ref{fig:2}}) with $0.4\text{ Mm}>z>-40\text{ Mm}$ and are scaled with $\sqrt{\rho}$.
					The initial wave packet has traveled through most of the atmosphere in this snapshot.
					Note the bending of the wavefronts (especially for $v_\text{diff}$) at large depths. This shows how the initially propagating wave turns back up towards the surface. 
					The wave emergence in the lower right panel is delayed by approximately $16$ min, which is the time it takes the initial wave packet to reach the center of the box, where the flux tube is located.
					\label{fig:2_2}
				}
			\end{figure*}

			Waves emerging from the source will behave as fast (fully) acoustic mode waves, once they cross the $c_\text{A} = c_\text{s}$-layer, they can convert to the slow magneto-acoustic mode branch \citep{2007AN....328..286C}. Slow mode waves will show two properties here: propagation preferably along magnetic field lines, and their transverse nature. Hence, showing the horizontal component will make these waves visible. As denoted in the top right panel of Figure \hyperref[fig:2]{\ref{fig:2}} by black arrows, slow mode waves will start to trickle downwards along the flux tube, as long as they are within the $c_\text{A} = c_\text{s}$-layer (and upwards out of the simulation domain). Once they cross that layer again, they can convert back to fast acoustic mode waves. These will now start to return to the surface after crossing their respective inner turning point, which is visualized in Figure \hyperref[fig:2_2]{\ref{fig:2_2}} for $t=60$ min and can best be seen for $z<-5$ Mm in the top right panel. Finally, the returning waves form a circular wave pattern centered on the flux tube, as seen in the bottom right panel of Figure \hyperref[fig:2_2]{\ref{fig:2_2}}.
			
			To summarize: waves that cross the flux tube boundary get partially converted and travel downwards, along field lines. They convert back again and travel to the surface. This is especially interesting, since the subsurface properties of the flux tube (at least within the $c_\text{A} = c_\text{s}$-layer) will have an influence on the wave field that can be observed at the surface. Indeed we expect that the travel-time, due to time being spent subsurface, and the shape, due to interference on the surface of the flux tube boundary, of the re-emerging wave packet will change. This change is thus related to the topology of the $c_\text{A} = c_\text{s}$-layer. A quantitative analysis is done in the next chapter.
			
			The surface signal $v_\text{diff}$ as seen in the bottom right of Figure \hyperref[fig:3]{\ref{fig:3}} is rather weak (about 10\% of the full wave field amplitude), but is still contained in $v_\text{magn}$ (as $v_\text{magn} = v_\text{diff} + v_\text{quiet}$). It is however superimposed by the wave packet of the original source, making it difficult to detect. In section \hyperref[sec:3]{\ref{sec:3}} \& \hyperref[sec:4]{\ref{sec:4}} it is shown how the two can be separated and potentially measured in real data.
			
			Using the difference $v_\text{diff}^\prime$ we observe a similar wave pattern, but with an amplitude of about 1\% of $v_\text{magn}$. From this behavior we learn that the re-emerging wave packet that is observed at the surface can not only stem from thermal modifications to the background atmosphere (due to the presence of a magnetic fluxtube), but must also carry contributions from the subsurface magnetic field itself. This is an important insight, meaning that $v_\text{diff}$ is in fact sensitive to subsurface magnetic fields.

\section{Time distance analysis with isolated sources}
\label{sec:3}
\begin{figure*}[htb]
		\centering
		\includegraphics[width=1.\textwidth]{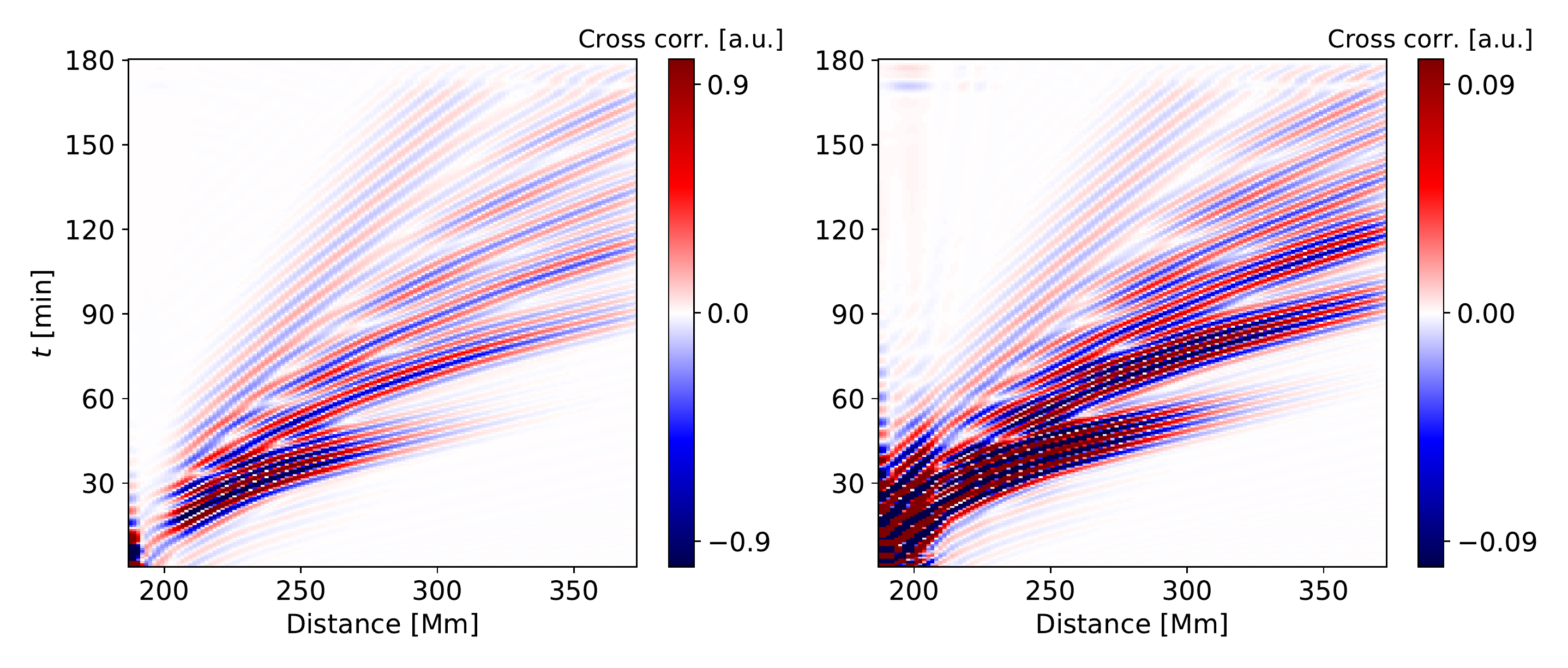}
		\caption{
			Time-distance diagrams for the single source analysis. Visible are branches of multiple skips for a frequency $\nu_\text{sp}$ of $3.3$ mHz for the full wave field $C_\text{magn}$ (left) and the instantaneous difference $C_\text{diff}$ (right) as described in eq. \hyperref[eq:Cdiff]{\ref{eq:Cdiff}}. Correlation is done for a point in the center of the flux tube (at 186.88 Mm) from the full simulation to an arc trailing outwards in the full simulation (left) and the subtracted simulation (right). Color therefore denotes correlation strength in arbitrary units. Note that the amplitude for $C_\text{diff}$ is about 10\% of the full wave field amplitude. No filter or averaging process is required for calculating the correlation, since only a single, isolated source is simulated. \label{fig:3}
			}
	\end{figure*}

	The method we use to describe the wave propagation quantitatively is the time-distance analysis \citep{1993Natur.362..430D, 2005LRSP....2....6G}. By cross-correlating the point $x_1$ in which the single source is located, with an arc $\bar{x}_2(\Delta)$ at different times $t$ and distances $\Delta$, we can measure quantities like the group travel-time $t_\text{g}$, the amplitude $A$ and the central frequency $\nu_0$ of the propagating wave packet. For this analysis the source is now put in the center of the simulation box $\vec{x_0} = \left( -0.4, 186.88, 186.88 \right)$ Mm, and thus $x_1$ is put directly within the flux tube. This positioning eliminates the initial travel time delay, meaning that the wave packet will instantly interact with the magnetic field, making the analysis more simple. The cross-correlation $C(t, \Delta)$ calculated as a function of $\Delta$ is then shown in Figure \hyperref[fig:3]{\ref{fig:3}}. For the two simulation runs $v_\text{magn}$ and $v_\text{quiet}$ we define:
	\begin{align} 
		\nonumber C_\text{magn}(t, \Delta) &\equiv \text{Cross-correlation for full magnetic run}\\
		\nonumber C_\text{quiet}(t, \Delta) &\equiv \text{Cross-correlation without magnetic field}\\ 
		C_\text{diff}(t, \Delta) &\equiv \text{Cross-correlation of the difference}
		\label{eq:Cdiff}
	\end{align}
	where by difference, again the instantaneous difference $v_\text{diff} = v_\text{magn} - v_\text{quiet}$ is meant. Note that the initial point $x_1$ for the cross-correlation $C_\text{diff}$ is taken from the full wave field $v_\text{magn}$, since we want to analyze the correlation of the waves contained in $v_\text{quiet}$ with those of the initial displacement. Since the amplitude of $v_\text{diff}$ is only about 10\% of that of $v_\text{magn}$, it is expected that this is also the case for the amplitude of $C_\text{diff}$, as can be seen in the right panel of Figure \hyperref[fig:3]{\ref{fig:3}}.

	The broad contributions that can be seen for $t > 90$ min and $\Delta < 210$ Mm can be assigned to a (unphysical but weak) reflection at the bottom boundary. Since these are restricted to an area that is generally not of interest, they can be ignored. Furthermore, the usual multiple skip branches are seen very clearly in both cases.

	Again, we repeat this analysis with the subtraction of thermal runs, i.e. $v_\text{therm}^\prime$, in which we replaced the difference signal accordingly to calculate the correlation $C_\text{diff}$. As expected, the main distinction is the amplitude. Due to the similarity of both correlation functions, we qualitatively conclude that $C_\text{diff}$ is in fact sensitive to the magnetic field.

\begin{figure*}[htb]
		\centering
		\includegraphics[width=1.\textwidth]{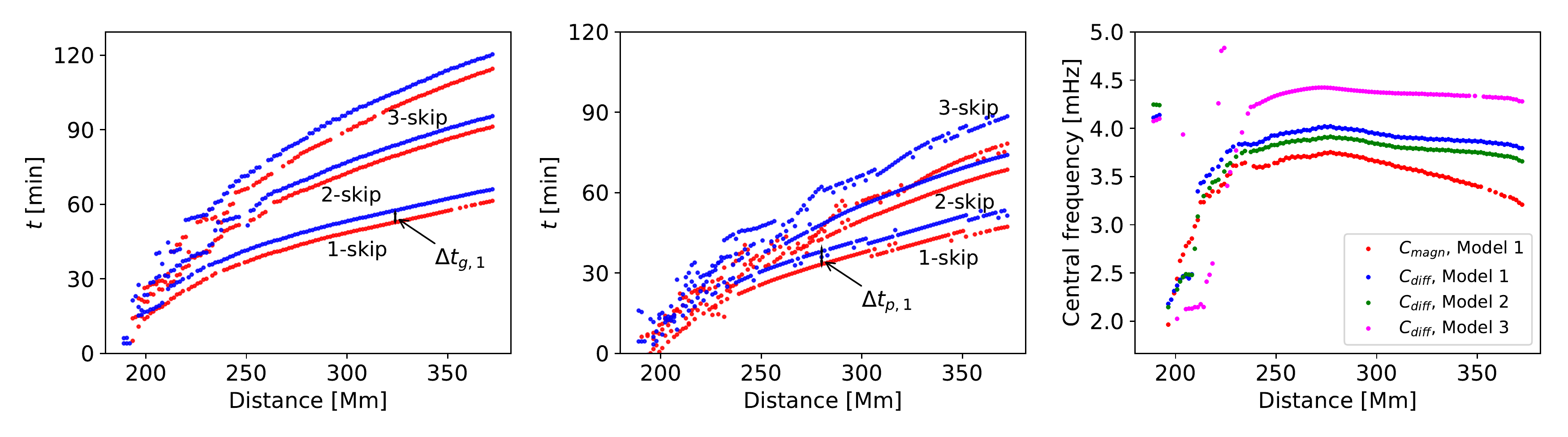}
		\caption{
			Results of fitting Gaussian wavelets to the different time-distance diagrams (as seen in Fig. \hyperref[fig:3]{\ref{fig:3}}). Left: Group travel times for the first three branches, for the full wave field $t_\text{g, magn}$ (red) and the instantenous difference of the regular monolithic flux tube model $t_\text{g, diff}$ (blue). Highlighted is also the very noticeable time delay of the group travel-times for the first branch $\varDelta t_{\text{g}, 1}$. Center: Phase travel times, also with the time delay $\varDelta t_\text{p, 1}$. Right: Central frequency $\nu_0$ in mHz of the fitted wavelet as a function of distance, for the full wave field (red), the regular monolithic flux tube model (model 1, blue), the alternative flux tube configuration (model 2, green) and the strongly modified model (model 3, magenta). \label{fig:4}
		}
	\end{figure*}

	As described in section \hyperref[sec:2.3]{\ref{sec:2.3}}, the wave packet caused by the oscillatory background displacement will travel a certain distance (depending on the depth of the inner turning point) before re-emerging at the surface. It is therefore expected that the group travel-times $t_\text{g, magn}$ of $C_\text{magn}$ and $t_\text{g, diff}$ of $C_\text{diff}$ differ. This can be measured by employing a fit to the data, to quantify the group travel-times separately. For this, Gabor-wavelets \citep{2005LRSP....2....6G} of the form 
	\begin{align}
		C(t) = A \exp\left( -(2\pi \text{d}\nu)^2 (t - t_\text{g})^2 \right) \cos\left( 2\pi\nu_0(t - t_\text{p}) \right)
		\label{eq:gbwl}
	\end{align}
	are used. With $A$ being the amplitude, $\text{d}\nu$ the width and $\nu_0$ the central frequency of the spectral distribution of the wavelet, $t_\text{g}$ the group speed and $t_\text{p}$ the phase speed. The results of $t_\text{g, magn}$ and $t_\text{g, diff}$ for this fit are shown for the first three branches in the left panel of Figure \hyperref[fig:4]{\ref{fig:4}}. A shift $\varDelta t_\text{g} = t_\text{g, magn} - t_\text{g, diff}$ can clearly be seen for all branches and is highlighted with an annotation for the 1-skip branch. Fitting the branches becomes more difficult starting from 3-skip, since for such distances, multiple skips start to overlap. Since $\varDelta t_\text{g, 1}$ only varies slightly with the distance $\Delta$, we estimate it by taking the average over $\Delta$. This yields:
	\begin{align}
		\varDelta t_\text{g, 1} = 282.6 \text{ sec}
		\label{eq:dtg}
	\end{align}
	The same analysis for $\varDelta t_\text{g, 2}$ and $\varDelta t_\text{g, 3}$ gives similar results, but is less precise, since the fit does not converge as consistently and is generally less robust. We conclude that $\varDelta t_\text{g} \equiv \varDelta t_\text{g, 1}$ is the most accurate estimation of the time delay.
	
	Shown in the center panels is the phase travel time $t_\text{p}$. Here a similar behavior is observed, although the time delay 
	\begin{align}
		\varDelta t_\text{p, 1} = 311.1 \text{ sec}
		\label{eq:dtp}
	\end{align}
	is more than twice as large as $t_\text{g, 1}$. However, the fit is much more erratic and becomes less reliable for small and large distances. For the analysis presented in the next section, we will be using $t_\text{g, 1}$ only, in principle it can nevertheless be performed with any quantity yielded by the fit.

	Since one of the goals of this work is learning about the subsurface configuration of the simulated flux tube, this study is repeated for the other two flux tube models (see Fig. \hyperref[fig:1.2]{\ref{fig:1.2}}). It turns out that changing the model affects $\varDelta t_\text{g}$ only weakly. However as shown in the right panel of Figure \hyperref[fig:4]{\ref{fig:4}} other parameters like the central frequency $\nu_0$ (see eq. \hyperref[eq:gbwl]{\ref{eq:gbwl}}) of the wavelet will show significant changes $\varDelta\nu_0$. While small changes in the model (model 1 to model 2) almost don't affect $\nu_0$ ($\varDelta\nu_0\approx0.04$ mHz), more drastic changes (model 1 to model 3) may increase the frequency by up to $\varDelta\nu_0\approx0.5$ mHz. This will allow to distinguish different subsurface magnetic field configurations, as long as the signal $C_\text{diff}$ is available, or can be reconstructed reasonably well (more on this in section \hyperref[sec:4]{\ref{sec:4}}).
	
	The fact that the group travel-time delay $\varDelta t_\text{g}$ is similar for all three models can be explained by considering the radius of the $c_\text{A} = c_\text{s}$-layer. Wave emergence visible in $v_\text{diff}$ will start at the boundary of said layer, leading to similar subsurface travel-times for similar radii $r(z)$. When slow mode waves are converted back to fast mode acoustic waves at the boundary, interference between them happens, depending on the shape of the $c_\text{A} = c_\text{s}$-layer. This interference will influence the wave pattern observed at the surface. That explains why $\nu_0(\Delta)$ (being an indicator for the spatial frequency-distribution of the wave packet) is very similar for model 2 and the original model 1, but shows significant changes of up to $0.5$ mHz for model 3, when the $c_\text{A} = c_\text{s}$-layer shape is changed drastically.
	
\section{Analysis with more realistic simulations}
\label{sec:4}
	Using real data, the instantaneous difference $v_{\text{diff}}$ is generally not available. 
	Any surface waves, that were caused by the mode conversion mechanism as described in \hyperref[sec:2.3]{\ref{sec:2.3}} (seen in the right panels of Fig. \hyperref[fig:2_2]{\ref{fig:2_2}}) are expected to be a feature of real sunspots as well. They will however have a small amplitude compared to other acoustic signals and will be swamped by them.
	In this section we will show one possible way to achieve a separation of the desired signal from the background nonetheless. 

	For this we ran two additional simulations with a solar like stochastic forcing function, similar to the one described in \cite{2007PhDT........24H} again with and without magnetic field. Since the simulation does not include radiative transfer and the background is artificially stabilized, the resulting velocity fields will not exhibit granulation. Thus the simulation loses some degree of realism. For most acoustic analyses however, granulation noise is unwanted anyway. The results are therefore treated as if granulation has been removed beforehand (by for example averaging or filtering processes). Looking again at the instantaneous difference $v_\text{diff}$, we see a circularly shaped wave pattern emerging from around the flux tube, equivalently to the behavior observed in Figure \hyperref[fig:2_2]{\ref{fig:2_2}}. Note that this is also briefly described in \cite{2007PhDT........24H}.

	We employ a time-distance analysis, similar to what we did in section \hyperref[sec:3]{\ref{sec:3}}, considering the fact that the cross-correlation signal originating from within the flux tube $C_\text{diff}$ is delayed by the time $\varDelta t_\text{g}$. This time however, the stochastic sources add a lot of unwanted contributions and noise. To still be able to construct a time-distance diagram, we employ a filtering process described in \cite{2005LRSP....2....6G} and rely on a point-to-arc average. A phase-speed filter that prefers waves with an average (one skip-) travel distance of $14.5$ Mm was applied. It is preferable to consider waves emerging close to the flux tube, since the amplitude of $v_\text{diff}$ is higher, due to its circular nature. Furthermore, we use points in a circle with a radius of $r_\text{C} = 29.31$ Mm around the flux tube (instead of directly inside it) to be correlated. Altogether, 616 point-to-arc cross-correlations functions are averaged.

\begin{figure*}[h!]
		\centering
		\includegraphics[width=1.\textwidth]{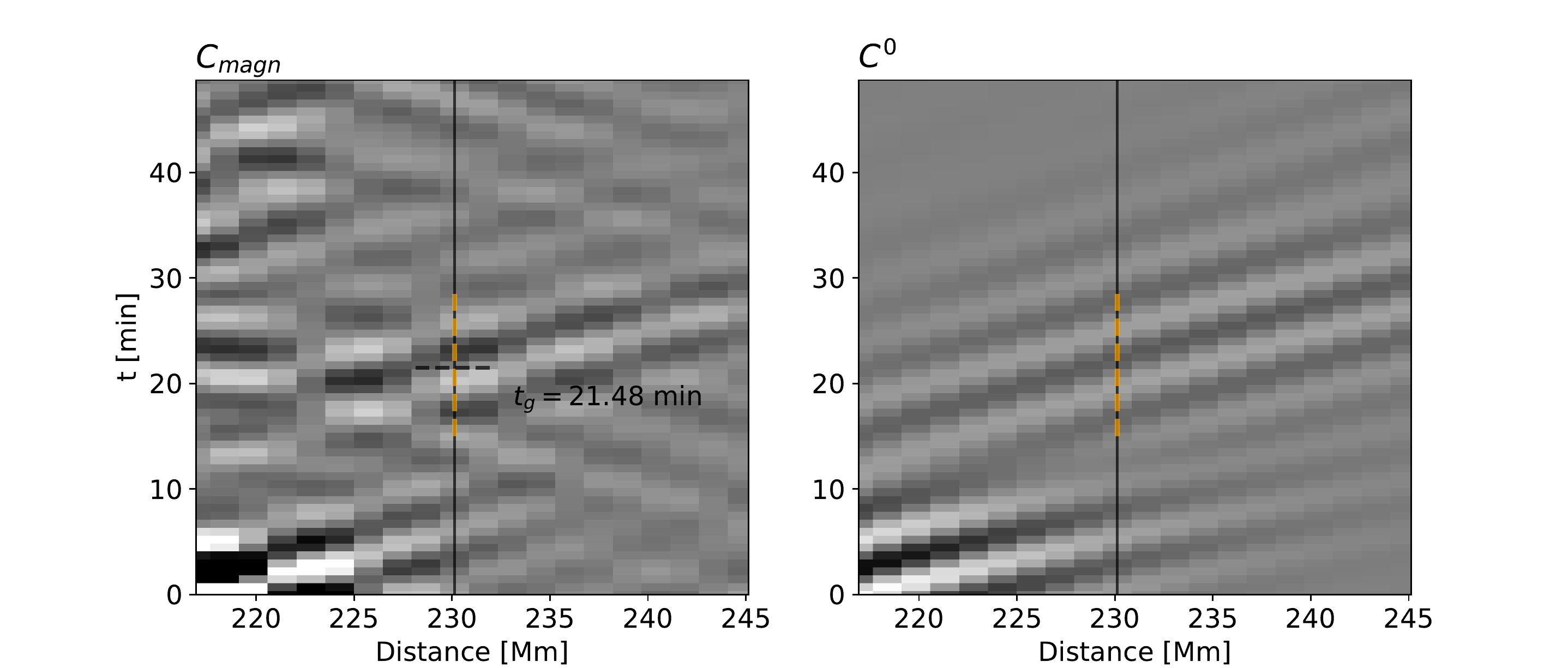}
		\caption{Cross-correlations of the full, multiple source wave field $C_\text{magn}$ (left) and the full, single source wave field (for the reference function) $C^0$ (right) as a function of time $t$ and distance. For both time-distance diagrams, a phase-speed filter was applied. For the left panel, an additional averaging process was applied. The right image appears smoother, since only a single source was simulated. The vertical black line marks the slice at which the fit was done, in this case $\Delta = 13.20$ Mm (distance $d = 230.13$ Mm). The dashed orange line indicates the width of the window function. The dashed horizontal line and annotation show the fit result. \label{fig:5_1}
			}
	\end{figure*}

	As mentioned, the flux tube signal will be weak and superimposed by the cross-correlation signal of the original wave packet $C_\text{magn}$, making it difficult to fit the desired signal $C_\text{diff}$ directly. The superposition will however modify the wavepacket in an asymmetric way, since it only contributes at specific times ($t + \varDelta t_g$, no contribution at $t - \varDelta t_g$). This can be quantified by using the \cite{2004ApJ...614..472G} fitting method. It requires essentially only a reference function $C^0(t, \Delta)$ that predicts the expected travel time at a certain distance $\Delta$. In our case this function is easily obtained by using single source simulations as in section \hyperref[sec:3]{\ref{sec:3}}. For real data one can either use theoretical predictions (i.e. the simulations done for this work, see Fig. \hyperref[fig:3]{\ref{fig:3}}), or a cross-correlation function from heavily averaged time-distance diagrams. The result of this fitting method is given in Figure \hyperref[fig:5_1]{\ref{fig:5_1}} at an exemplary distance $\Delta = 13.2$ Mm, where the travel-time $t_\text{g, magn}$ is annotated within the according image. A window with a width of $\sigma_\text{window} = 13.5$ min (shown as dashed orange line) was applied the interval of estimated travel-time.
	
	\begin{figure*}[h!]
		\centering
		\includegraphics[width=1.\textwidth]{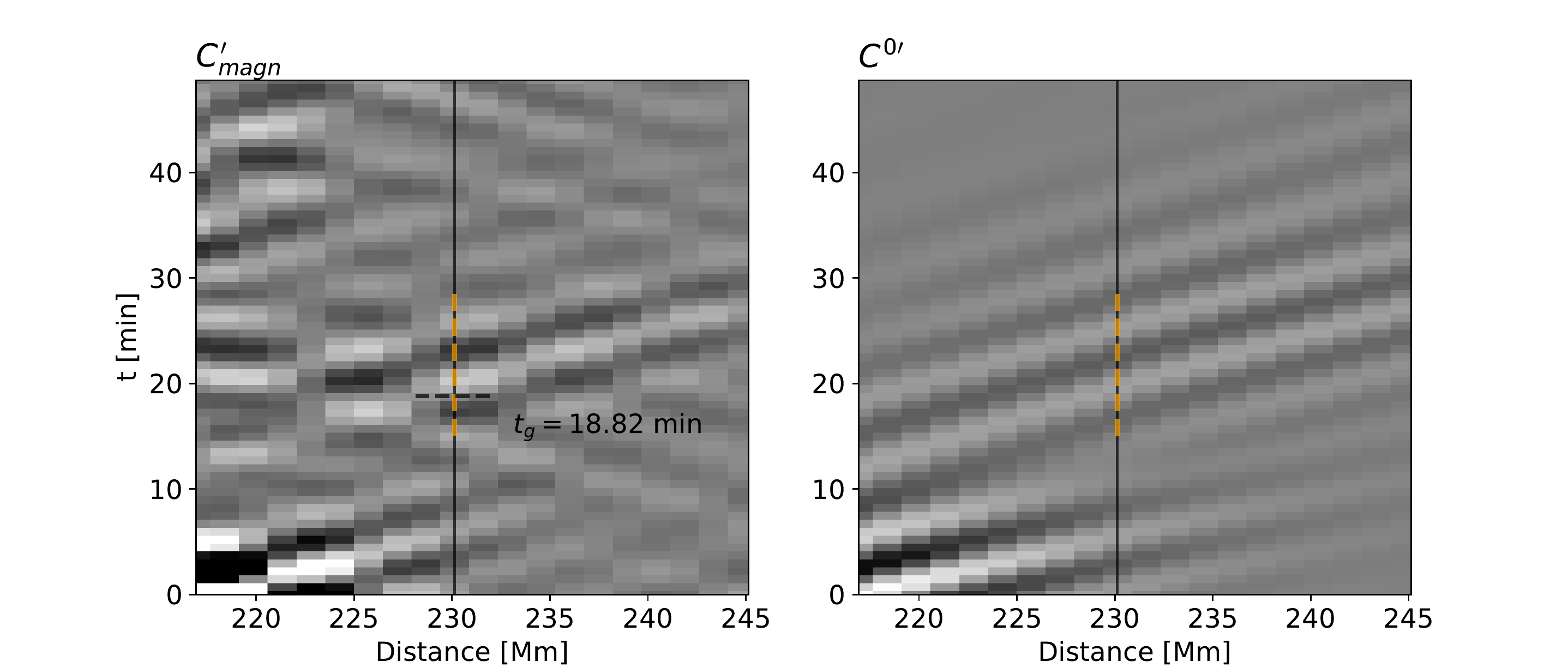}
		\caption{
			Same as Figure \hyperref[fig:5_1]{\ref{fig:5_1}}, except that primed quantities, i.e. time-distance diagrams where the correlation signal due to presence of the magnetic field has been removed, are shown (as in eq. \hyperref[eq:prime]{\ref{eq:prime}}). 
			The travel-time $t_\text{g}$ is decreased, due to the correction shown in equation \hyperref[eq:prime]{\ref{eq:prime}}.
			\label{fig:5_2}
		}
	\end{figure*}
\begin{figure}[h!]
		\plotone{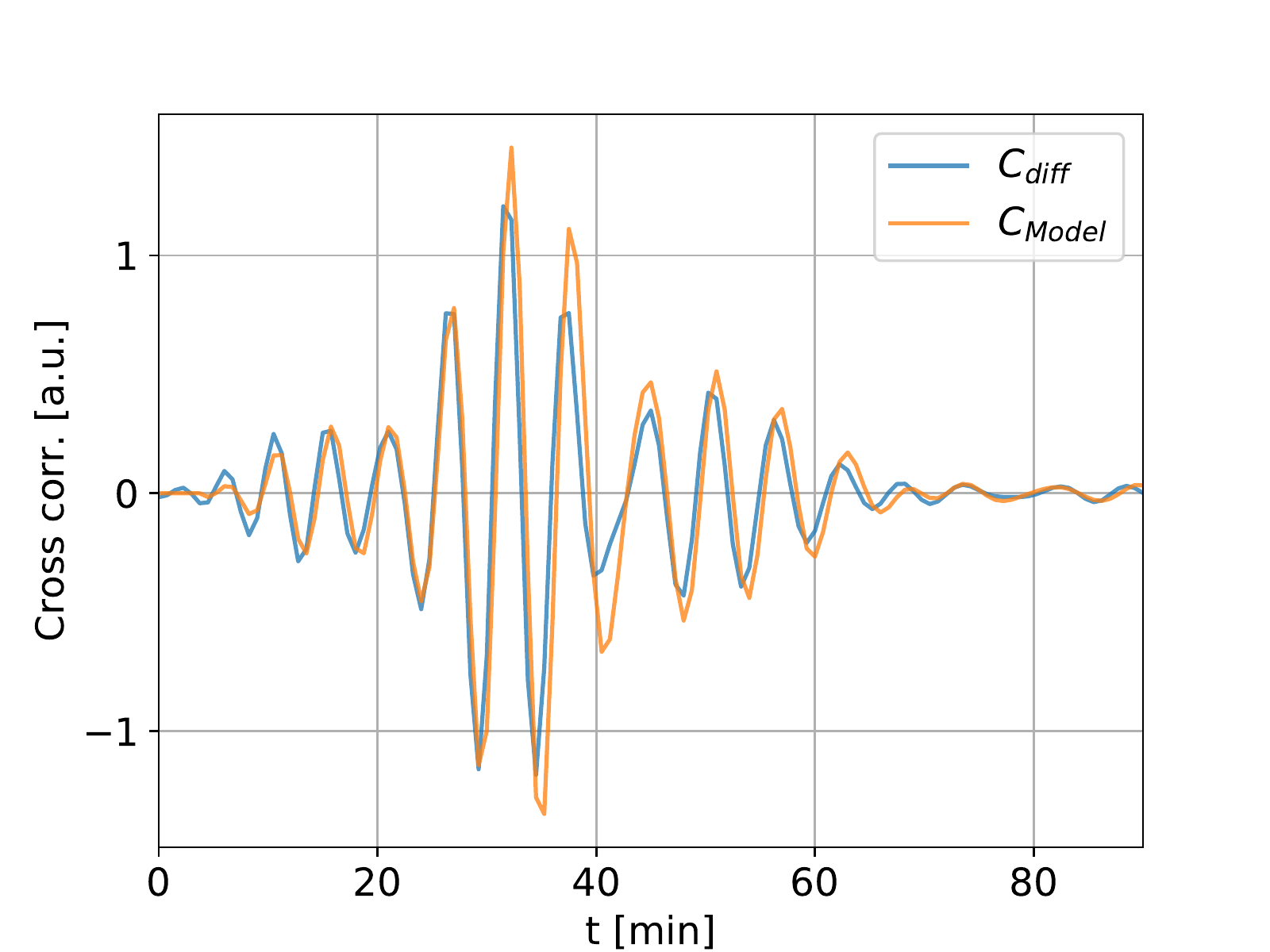}
		\caption{The desired (but in real data unknown) cross-correlation signal of $C_\text{diff}$ (blue) as a function of time $t$ at a distance of $\Delta = 29.31$ Mm (distance $d = 246.24$ Mm). Also plotted is the model attempt $C_\text{Model}$ (orange), constructed from the reference function $C^0$ (see eq. \hyperref[eq:mod]{\ref{eq:mod}}). \label{fig:5_3}}
	\end{figure}

	The challenge is now to detect a change in the travel-time due to the emerging wave signal of the flux tube. We do this by \textquotedblleft correcting\textquotedblright\space the data $C_\text{magn}(t, \Delta)$, and the reference function $C^0(t, \Delta)$. In principle the time-distance diagram on the right in Figure \hyperref[fig:5_1]{\ref{fig:5_1}} exhibits contributions as shown on the right in Figure \hyperref[fig:3]{\ref{fig:3}}. This can be modeled, as long as we can estimate $t_\text{g, diff}$ and the amplitude $A_\text{diff}$ of the cross-correlation $C_\text{diff}$ with $t_{\text{g, Model}}$ and $A_\text{Model}$. The actual cross-correlation functions for the full wavefield $C_\text{magn}$ and instantaneous difference $C_\text{diff}$ differ in more than just the amplitude and travel-time (see Fig. \hyperref[fig:4]{\ref{fig:4}}, right panel), which is however neglected for the simplicity of this analysis. For the correction, we simply subtract the contribution of the flux tube:
	\begin{align}
		C_\text{magn}^\prime &= C_\text{magn} - C_\text{Model}\\
		C^{0\prime} &= C^0 - C_\text{Model}
		\label{eq:prime}
	\end{align}
	where primed quantities means corrected. 
	Correcting the data like this assumes that the difference signal $v_\text{diff}$ contributes to the Correlation $C_\text{magn}$ in a linear fashion. Generally the construction of $C_\text{magn}$ from $v_\text{magn}$ and $v_\text{diff}$ is more complicated, therefore the correction in equation \hyperref[eq:prime]{\ref{eq:prime}} is based on a first order approximation.
	$C_\text{Model}$ is then constructed via 
	\begin{align}
		C_\text{Model}(t, \Delta) = A_\text{Model} \cdot C^0\left( t + \varDelta t_\text{g}, \Delta \right)\text{  ,}
		\label{eq:mod}
	\end{align}
	with, in this case $\varDelta t_\text{g} = t_\text{g, Model} - t^0_\text{g}$.
	An exemplary attempt at constructing the model $C_\text{Model}$ is shown in Figure \hyperref[fig:5_3]{\ref{fig:5_3}}. Fitting the corrected data $C_\text{magn}^\prime$ will then yield a slightly modified travel-time $t_\text{g}^\prime$, shown in Figure \hyperref[fig:5_2]{\ref{fig:5_2}}. The difference for the two results $t_\text{g} - t_\text{g}^\prime$ is not equal to $\varDelta t_\text{g}$, as it is only a slight modification due to the contribution of the flux tube signal $C_\text{diff}$. It is however related to $\varDelta t_\text{g}$, by how well the model attempt $C_\text{Model}$ agrees with $C_\text{diff}$. Generally it is expected that $t_\text{g}^\prime < t_\text{g}$, since $t_\text{g, magn} > t_\text{g, quiet}$. Also, if $C_\text{Model}$ does not agree well with $C_\text{diff}$, it is expected that $t_\text{g} - t_\text{g}^\prime < \varDelta t_\text{g}$. Summarizing, we can make the first constraint on the measurement of $t_\text{g}^\prime$, and therefore for the agreement between $C_\text{Model}$ and $C_\text{diff}$:
	\begin{align}
		t_\text{g} - \varDelta t_\text{g} < t_\text{g}^\prime < t_\text{g}\text{  .}
		\label{eq:4.1}
	\end{align}
	In our exemplary analysis, we find $t_\text{g} - t_\text{g}^\prime = 159.7$ sec, where we set ($t_\text{g, Model} - t^0_\text{g} = $) $\varDelta t_g = 270.0$ sec and $A_\text{Model} = 0.20$. Since for simulations, $C_\text{diff}$ is available, we can calculate the theoretical value of $t_\text{g}^\prime$ and deduce $\varDelta t_g$ from fits, as shown in Figure \hyperref[fig:4]{\ref{fig:4}}, for comparison. Here we find:
	\begin{align}
		t_\text{g} - t_\text{g}^\prime &= 191.7 \text{ sec}\\
		\varDelta t_g &= 282.6 \text{ sec}\text{  .}
	\end{align}
	As can be seen, the initial estimate of $\varDelta t_g = 270.0$ sec leads to a $t_\text{g} - t_\text{g}^\prime$ that is already close to the theoretical value.

	The constraint \hyperref[eq:4.1]{\ref{eq:4.1}} yields a broad estimate on how to choose the parameters $A_\text{Model}$ and $t_\text{g, Model}$ for the model attempt $C_\text{Model}$, but the exact value of $\varDelta t_\text{g}$ will remain unknown for real data. One would need to do this analysis for several sunspots, to further narrow down eq. \hyperref[eq:4.1]{\ref{eq:4.1}}, or rely on the simulations done in this work, to have a reference for the required estimate of $\varDelta t_\text{g}$.\\

\section{Discussion}
	The mechanism of acoustic fast mode waves being converted into downward propagating slow mode waves is a known process, \citep{2003MNRAS.346..381C, 2015ApJ...801...27R, 2016ApJ...817...45R} but has however never been studied in particular. Moreover the ramping effect \citep{2007AN....328..286C}, which is similar in its nature (upward propagation instead of the downward propagation considered here), was investigated more extensively, due to its possible contribution to acoustic halos. In this work, it was shown that flux tubes behave as sources of acoustic power, as long as they are being excited from the outside (sec. \hyperref[sec:2.3]{\ref{sec:2.3}}). The consequence is that emerging waves from a sunspot contribute to the acoustic power in its vicinity. Of course the absorption of p-mode power \citep{1987ApJ...319L..27B, 1988ApJ...335.1015B}, makes detecting this power excess in real data non-trivial.
	
	The time delay $\varDelta t_\text{g}$ shown in Figure \hyperref[fig:4]{\ref{fig:4}} is a way to distinguish the signals $C_\text{magn}$ and $C_\text{diff}$. Setting up simulations with solar like sources, it was shown that although the amplitude of the $C_\text{diff}$ is low, $\varDelta t_\text{g}$ can still be estimated. As shown in section \hyperref[sec:3]{\ref{sec:3}} (left panel of Fig. \hyperref[fig:4]{\ref{fig:4}}), a decent reconstruction of $C_\text{diff}$ via $C_\text{Model}$ using an estimate for $\varDelta t_\text{g}$ may reveal subsurface properties of the investigated flux tube. It is necessary however to obtain the wave pattern of $C_\text{diff}$ to get d$\nu$ and $\nu_0$ (see eq. \hyperref[eq:gbwl]{\ref{eq:gbwl}}). Since $t_\text{g}$ does not vary for different flux tube models, $d\nu$ and $\nu_0$ are needed to relate the surface signal to subsurface properties. It also appears that slight changes to the model that do not affect the $c_\text{A} = c_\text{s}$-layer do not affect the surface signal $C_\text{diff}$. For real data this will require additional effort in creating $C_\text{Model}$.

	With the proposed method in section \hyperref[sec:4]{\ref{sec:4}}, a measurement of $t_\text{g}^\prime$ needs to be done reliably, which will be difficult in the case of real data. In principle, the method can be done for more sets of filters , as described in \cite{2005LRSP....2....6G} and thus, more distances $\Delta$. We tested this here, and got similar results, but not for all distances $\Delta$. This is expected, since, as mentioned, the amplitude of $C_\text{diff}$ becomes weaker for large $\Delta$, decreasing the signal-to-noise ratio, due to its circular wave behavior. Again, another hurdle in estimating $\varDelta t_g$ reliably. 

\section{Conclusion}
	Using the SPARC code to simulate the interaction of different kinds of waves with simple flux tube models, the effects of mode conversion have been visualized (see Fig. \hyperref[fig:2]{\ref{fig:2}}). Slow mode waves traveling downwards along magnetic field lines of the flux tube convert back to acoustic fast mode waves, that deflect back up and are measurable at the surface. It is demonstrated how these surface waves are altered from subsurface changes in the flux tube model (see Fig. \hyperref[fig:3]{\ref{fig:3}}). Moreover, the fact that these waves spend time traveling within the flux tube, a time delay $\varDelta t_\text{g}$ (see eq. \hyperref[eq:dtg]{\ref{eq:dtg}}) between re-emerging wave and original (as caused by the initial source) wave can be measured. We find that $\varDelta t_\text{g} = 282.6$ sec. It was also found that subsurface changes in the flux tube models, especially changes to the shape of the $c_\text{A} = c_\text{s}$-layer, influence the frequency distribution of the surface wave pattern (see Fig. \hyperref[fig:4]{\ref{fig:4}}). 
	
	A method to estimate $\varDelta t_\text{g}$ for real data is presented in section \hyperref[sec:4]{\ref{sec:4}}. Although the method might become unreliable for real data due to many sources of noise, for our simulations a value of $t_\text{g} - t_\text{g}^\prime = 159.7$ sec with an assumption of $\varDelta t_\text{g} = 225$ sec proved to be accurate when compared to the theoretically predicted value of $t_\text{g} - t_\text{g}^\prime = 191.7 \text{ sec}$ and the associated $\varDelta t_g = 282.6 \text{ sec}$.
	
	The reconstruction of $C_\text{Model}$ with quantities available in real data, in order to estimate $C_\text{diff}$ will need some additional effort, to be reliable. Also, measuring properties of the wave pattern, like the central frequency $\nu_0$ might require direct detection of $C_\text{diff}$, which again, will be difficult due to its comparatively low amplitude. 
	
	This work serves as a theoretical basis for a new method with the potential of adding to the knowledge of subsurface sunspot properties.
	The next step for further analysis regarding this topic is executing this study for real data and making these simulations more realistic by for example tuning the flux tube model, the background model etc. This will include fine tuning the proposed method as for example equation \hyperref[eq:4.1]{\ref{eq:4.1}}.
	
	\textit{Acknowledgments}: We thank Shravan M. Hanasoge for making the SPARC code, being the basis for this work, publicly available at \url{http://www2.mps.mpg.de/projects/seismo/sparc/}.
	
\bibliographystyle{aasjournal}

\end{document}